\documentstyle[floats,prl,aps]{revtex} 
\input epsf

\def\spose#1{\hbox to 0pt{#1\hss}}
\def\simlt{\mathrel{\spose{\lower 3pt\hbox{$\mathchar"218$}}
        \raise 2.0pt\hbox{$\mathchar"13C$}}}
\def\simgt{\mathrel{\spose{\lower 3pt\hbox{$\mathchar"218$}}
     \raise 2.0pt\hbox{$\mathchar"13E$}}}

\begin{document}

\def\degrees{\hbox{${}^\circ$\hskip-3pt .}}
\def\spose#1{\hbox to 0pt{#1\hss}}
\def\simlt{\mathrel{\spose{\lower 3pt\hbox{$\mathchar"218$}}
        \raise 2.0pt\hbox{$\mathchar"13C$}}}

\newcommand{\Phigb}{\Phi_{\gamma b}}
\newcommand{\RF}{{\cal{T}}}
\newcommand{\Aa}{{\cal A}_a}
\newcommand{\Ab}{{\cal A}_b}
\newcommand{\bg}{{b\gamma}}
\newcommand{\eal}{\!\!\! & = & \!\!\!}

\preprint{IASSNS-96/7}
\title{A NEW TEST OF INFLATION}
\author{Wayne Hu${}^1$ \& Martin White${}^2$}
\address{${}^1$Institute for Advanced Study, School of Natural Sciences\\
Princeton, NJ 08540 \\
${}^2$Enrico Fermi Institute, University of Chicago\\
Chicago, IL 60637}

\twocolumn[
\maketitle
\widetext

\begin{abstract}We 
discuss a new test of inflation from the harmonic pattern of peaks
in the cosmic microwave background radiation angular power spectrum.
By characterizing the features of alternate models and revealing
signatures essentially
unique to inflation, we show that inflation could be validated
by the next generation of experiments. 
\end{abstract}
\hskip 0.5truecm
]
\narrowtext


Inflation is the front running candidate for generating fluctuations in the
early universe: the density perturbations which are the precursors of galaxies
and cosmic microwave background (CMB) anisotropies today.
By ``inflation'' in this letter we shall mean simply the idea that the
universe underwent a period of vacuum driven superluminal expansion during
its early evolution, which provides a mechanism of connecting, at early times,
parts of the universe which are currently space-like separated.
It has been argued that inflation is the unique {\it causal} mechanism for
generating correlated curvature perturbations on scales larger than the
horizon \cite{acausal,BigPaper}.
If there are unique consequences of such super-horizon curvature perturbations,
their observation would provide strong evidence for inflation.

\begin{figure}[t]
\begin{center}
\leavevmode
\epsfxsize=3.5in \epsfbox{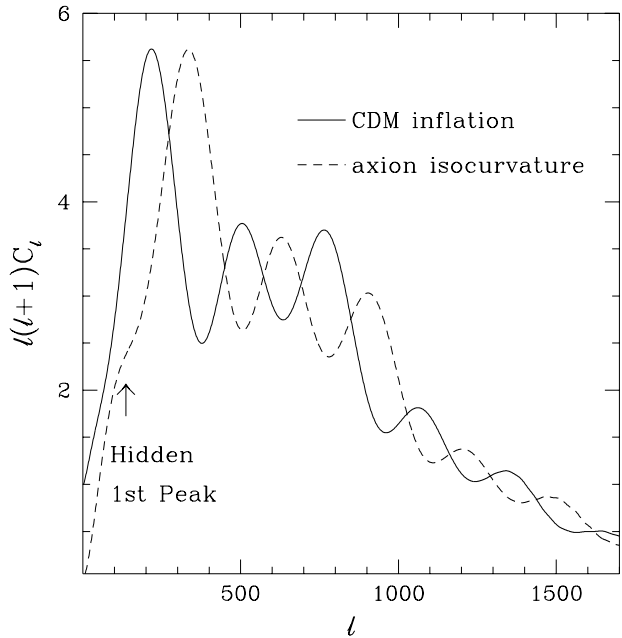}
\end{center}
\caption{\noindent
The angular power spectrum of a ``standard'' inflationary
CDM model with $\Omega_0=1$, $h=0.5$ and $\Omega_b=0.05$ (solid).
The $x$-axis is multipole number $\ell\sim\theta^{-1}$.  The peaks on angular
scales below $0\degrees5$ ($\ell\simgt200$) are the result of acoustic
oscillations in the photon-baryon fluid before recombination.
Also shown (dashed) is the spectrum for an axionic isocurvature model with
the same cosmological parameters.
Note the peaks are offset from the inflationary prediction, and the first
``peak'' is more of a shoulder in this model.}
\label{fig:scdm}
\end{figure}

In this {\it Letter}, we probe the nature of the fluctuations through CMB
anisotropy observations of the acoustic signatures in the spectrum.
Many of the relevant technical details as well as more subtle examples can
be found in \cite{BigPaper}.
As a working hypothesis, we shall assume that the CMB spectrum exhibits a
significant harmonic signature: a series of peaks in the power spectrum when
plotted against multipole number $\ell$ (see Fig.~\ref{fig:scdm}; for reviews
of the underlying physics of these peaks see \cite{HuSug,HSb,acoustic}).
Such a signature is expected in inflationary models and is characterized by
the locations and relative heights of the peaks as well as the position of
the damping tail.
We will comment at the end on situations where inflation may have happened
without leaving us this clear signature. 

The possibility of distinguishing some specific defect models from inflation
based on the structure of the power spectrum below $0\degrees5$ has recently
been emphasized \cite{CriTur,Albetal}.
By characterizing the features of such alternate models and revealing
signatures unique to inflation, here and in \cite{BigPaper}, we provide the
extra ingredients necessary to allow a test of the inflationary paradigm.
Another means of testing inflation is the consistency relation between the
ratio of tensor and scalar modes and the tensor spectral index
\cite{Davetal,KnoTur}.  However this test requires a large tensor signal
\cite{KnoTur} or it will be lost in the cosmic variance.


In Fig.~\ref{fig:scdm} (solid line), we show the angular power spectrum of
CMB anisotropies for a standard cold dark matter (CDM) inflationary model,
as a function of multipole number $\ell\sim\theta^{-1}$.
To understand the features in the spectrum below $0\degrees5$
($\ell\simgt200$), consider the universe just before it cooled enough to allow
protons to capture electrons.
At these early times, the photons and baryon-electron plasma are tightly
coupled by Compton scattering and electromagnetic interactions.
These components thus behaved as a single `photon-baryon fluid' with the
photons providing the pressure and the baryons providing inertia.
In the presence of a gravitational potential, forced acoustic oscillations in
the photon-baryon fluid arise.
The energy density, or brightness, fluctuations in the photons are seen by the
observers as temperature anisotropies on the CMB sky.
Specifically, if $\Theta_0$ is the temperature fluctuation $\Delta T/T$ in 
normal mode $k$, the oscillator equation is
\begin{equation}
{d \over d\eta}\left[ m_{\rm eff}{d\Theta_0\over d\eta}\right] +
  {k^2 \over 3}\Theta_0
  = -F[\Phi,\Psi,R]
\label{eqn:Oscillator}
\end{equation}
with
\begin{equation}
F[\Phi,\Psi,R]= {k^2 \over 3}m_{\rm eff}\Psi +
  {d \over d\eta}\left[m_{\rm eff}{d\Phi\over d\eta}\right],
\end{equation}
where $m_{\rm eff}=1+R$, $R=3\rho_b/4\rho_\gamma$ is the baryon-to-photon
momentum density ratio, $\eta=\int dt/a$ is conformal time,
$\Phi$ is the Newtonian curvature perturbation, and 
$\Psi \approx -\Phi$ is the gravitational potential \cite{BigPaper,HuSug,HSb}.

In an inflationary model, the curvature or potential fluctuations are created
at very early times and remain constant until the fluctuation crosses the
sound horizon.  
As a function of time, this force excites a cosine mode of the acoustic
oscillation.  The first feature represents a compression of the fluid inside
the potential well as will become important in the discussion below.
Furthermore, the harmonic series of acoustic peak location
$\ell_1:\ell_2:\ell_3 $ approximately follows the cosine series of extrema
$1:2:3\cdots$ \cite{SeriesNote}.
There are two concerns that need to be addressed for this potential test of
inflation.  How robust is the harmonic prediction in the general class of
inflationary models? Can any other model mimic the inflationary series?  

\begin{figure}[t]
\begin{center}
\leavevmode
\epsfxsize=3.5in \epsfbox{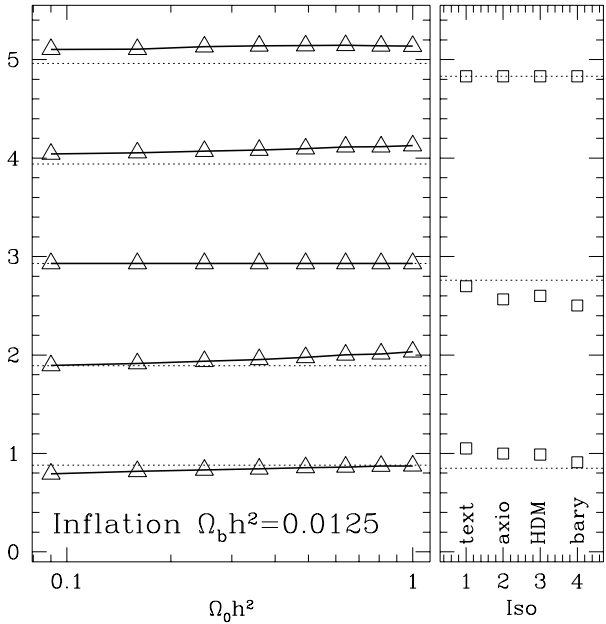}
\end{center}
\caption{\noindent 
The relative positions of the peaks in the angular power spectrum
$\ell_1 : \ell_2 : \ell_3 \cdots $ for the inflationary 
(left panel, points) and 4 isocurvature models (right panel, points):
textures; axion isocurvature; hot dark matter isocurvature and
baryon isocurvature (see text).  
The series are normalized at $\ell_3$ to the idealized inflationary
and isocurvature series respectively (dotted line) [10]. 
Test cases illustrate that the two cases remain
quite distinct, especially in the ratio of the first to third peak and
peak spacing.}
\label{fig:peak}
\end{figure}

The peak ratios are not affected by the presence of spatial curvature or
a cosmological constant in the universe \cite{BigPaper}.  However it is
possible to distort the shape of the first peak by non-trivial evolution of
the metric fluctuations after last scattering.   
For example, the magnitude of the scalar effect increases with the influence of
the radiation on the gravitational potentials, e.g.~by a decrease in the matter
content $\Omega_0h^2$.  Here $\Omega_0$ is the current matter density in units
of the critical density and the Hubble constant is
$H_0=100h$km s$^{-1}$ Mpc$^{-1}$.
Tensor fluctuations could distort the first peak and spectral tilt shift the
series only if they are very large.  That possibility is inconsistent with the
observed power at degree scales \cite{TiltNote}.
The damping of power in the oscillations at small scales due to photon
diffusion cuts off the spectrum of peaks and could also confuse a measurement
of their location.
In Fig.~\ref{fig:peak}, we plot the ratio of peak locations as a function of
$\Omega_0 h^2$.
Although the first peak is indeed slightly low in the low $\Omega_0h^2$ models
the harmonic series is still clearly discernible in the regular spacing of the
higher peaks.  Two numbers serve to quantify the spectrum: the
ratio of third to first peak location $\ell_3/\ell_1 \approx 3.3-3.7$
and the first peak location to the spacing between the peaks 
$\ell_1/\Delta \ell \approx 0.7-0.9$.  Ratios in this range are 
a robust prediction of inflation with reasonable baryon 
content \cite{BaryonNote}.

Is the cosine harmonic series a {\it unique} prediction of inflation?
Causality requires that all other models form significant 
curvature perturbations near or after horizon crossing \cite{IsocNote}.
We call these {\it isocurvature} models.  The axionic isocurvature model
of Fig.~1 (dashed lines) is representative \cite{KawSugYan}.  
Since curvature fluctuations
start out small and grow until horizon crossing, the peak locations
are phase shifted with respect to the inflationary prediction.  
In typical models, including 
the baryon isocurvature \cite{HSb},
texture \cite{CriTur}, 
axionic isocurvature \cite{KawSugYan}, 
hot dark matter isocurvature \cite{deLSch},
the peaks approximately form a sine series $1:3:5\cdots$ \cite{SeriesNote}
(see Fig.~2, right panel).
More generally, isocurvature models introduce some phase shift with respect to
the inflationary prediction.  The {\it spacing} of the peaks $\Delta\ell$
however remains unchanged as it reflects the natural period of the oscillator.

How might an isocurvature scenario mimic the inflationary prediction?
Two possibilities arise.  If the first isocurvature feature, which is
intrinsically low in amplitude is hidden, e.g.~by external metric fluctuations
such as tensor and vector contributions between last scattering and the
present, the series becomes approximately $3:5:7$.
Might this be mistaken for an inflationary spectrum, shifted to smaller angles
by the curvature of the universe?
The spectra remain distinct since the spacing between the peaks is fixed.
The ratio of the first peak position to peak spacing $\ell_1/\Delta\ell$ is
larger by a factor of $1.5$ in this case if $\Omega_0 h^2$ is fixed.
In \cite{BigPaper}, we treat the ambiguity that arises if this and other
background quantities are unknown.
More generally, any isocurvature model that either introduces a pure phase
shift or generates acoustic oscillations only well inside the causal horizon
can be distinguished by this test.  Of course, isocurvature models need not
exhibit a simple regularly-spaced series of peaks \cite{BigPaper,Albetal},
but these alternatives could not mimic inflation.


The remaining possibility is that an isocurvature model might be tuned so that
its phase shift precisely matches the inflationary prediction.
Heuristically, this moves the whole isocurvature spectrum in Fig.~1 toward
{\it smaller} angles.  We shall see that causality forbids us to make the
shift in the opposite direction.
As Fig.~1 implies, the relative peak heights can distinguish this possibility
from the inflationary case.

The important distinction comes from the process of compensation, required
by causality.
During the evolution of the universe, the dominant dynamical component
counters any change in the curvature produced by an arbitrary source
\cite{BigPaper}.
Producing a positive curvature perturbation locally stretches space.  The
density of the dominant dynamical component is thus reduced in this region,
and hence its energy density is also reduced.  This energy density however
contributes to the curvature of space, thus this reduction serves to offset
the increased curvature from the source.
Heuristically, curvature perturbations form only through the motion of
matter, which causality forbids above the horizon.  

In the standard scenario, the universe is radiation dominated when the smallest
scales enter the horizon (see \cite{BigPaper} for exotic models).
Thus near or above the horizon, the photons resist any change in curvature
introduced by the source.  Breaking $\Phi$ into pieces generated by the
photon-baryon fluid ($\gamma b$) and an external source ($s$), we find its
evolution in this limit follows \cite{BigPaper}
\begin{equation}
x^2 \Phigb'' + 4 x \Phigb' = - x^2 \Phi_s'' - 4 x \Phi_s' ,
\label{eqn:Compensationa}
\end{equation}
where primes denote derivatives with respect to $x = k\eta$.

Thus the first peak in an isocurvature model, if it is sufficiently close to
the horizon to be confused with the inflationary prediction and follows the
cosine series defined by the higher peaks, must have photon-baryon fluctuations
anti-correlated with the source.  
The first peak in the rms temperature thus represents the {\it rarefaction} (r)
stage when the source is overdense rather than a compression (c)
phase as in the inflationary prediction.  The peaks in the inflationary
spectrum obey a c-r-c pattern while the isocurvature model displays a r-c-r
pattern.  Though compressions and rarefactions have the same amount of
power (squared fluctuation), there is a physical effect which allows us to
distinguish the two: baryons provide extra inertia to the photons to which
they are tightly coupled by Compton scattering (the $m_{\rm eff}$ terms in
Eq.~\ref{eqn:Oscillator}).  If overdense regions represent potential
wells \cite{TurokNote}, this inertia enhances compressions at the 
expense of 
rarefactions leading to
an alternating series of peaks in the rms \cite{HuSug}.  For reasonable baryon
content \cite{BaryonNote}, the {\it even} peaks of an isocurvature model are
enhanced by the baryon content whereas the {\it odd} peaks are enhanced under
the inflationary paradigm (see Fig.~1).  
This is a non-monotonic modulation of the peaks so is not likely to occur in
the initial spectrum of fluctuations.
The oscillations could be driven at exactly the natural frequency of the
oscillator in such a way as to counteract this shift, but such a long duration
tuned driving seems contrived.

There is one important point to bear in mind.  Since photon diffusion damps
power on small scales, the 2nd compression (3rd peak) in an inflationary model
may not be {\it higher} than the 1st rarefaction (2nd peak), even though it is
enhanced (see e.g.~Fig.~\ref{fig:scdm}).
However it will still be anomalously high compared to a rarefaction peak,
which would be both suppressed by the baryons and damped by diffusion.
Since the damping is well understood this poses no problem in principle
\cite{BigPaper}.

Diffusion damping also supplies an important consistency test.
The physical scale depends only on the background cosmology and not on the
model for structure formation (see Fig.~1 and \cite{HuSug,BigPaper}),
\begin{equation}
k_D^{-2} = {1 \over 6} \int d\eta\ {1\over\dot{\tau}}\,
{R^2 + 16(1+R)/15 \over (1+R)^2},
\label{eqn:DiffusionLength}
\end{equation}
where $\dot{\tau}=n_e\sigma_Ta$ is the differential optical depth to Compton
scattering.  This fixed scale provides another measure of the phase shift
introduced by isocurvature models.  For example, if the first isocurvature peak
in Fig.~1 is hidden, the ratio of peak to damping scale increases by a factor
of 1.5 over the inflationary models.  We also consider in \cite{BigPaper} how
the damping scale may be used to test against exotic background parameters and
thermal histories.


In summary, the ratio of peak locations is a robust prediction of inflation.
If acoustic oscillations are observed in the CMB, and the ratio of the 3rd to
1st peak is not in the range $3.3-3.7$ or the 1st peak to peak-spacing in
the range $0.7-0.9$ then either inflation does not provide the main source of
perturbations in the early universe or big bang nucleosynthesis grossly
misestimates the baryon fraction \cite{BaryonNote}.
The ranges can be tightened if $\Omega_0 h^2$ is known.
If the spatial curvature of the universe vanishes, these tests require CMB
measurements from $10-30$ arcminutes.  Even if the location of the first peak
is ambiguous, as might be the case in some isocurvature models, these tests
distinguish them from inflation.
Isocurvature models thus require fine tuning to reproduce this spectrum.
To close this loophole, the relative peak heights can be observed.  Assuming
the location of the peaks follows the inflationary prediction, the enhancement
of odd peaks is a unique prediction of inflation \cite{TurokNote}.

As for failing to test inflation through these means, there are
several possible but unlikely scenarios.  Strong early reionization
$z\simgt100$ could erase the acoustic signature.
Inflation also does not preclude the presence of isocurvature 
perturbations, so their presence does not rule out inflation.

The true discriminatory power of the CMB manifests itself in the spectrum as
a whole, from degree scales into the damping region.
In particular, we emphasize the acoustic {\it pattern} which arises from
forced oscillations of the photon-baryon fluid before recombination, including 
the model-independent nature of the damping tail.
The tests we describe rely on the gross features of the angular power spectrum
and so could be performed with the upcoming generation of array receivers and
interferometers.

\bigskip
We thank 
{J. Bahcall},
{P.G. Ferreira},
A.~Kosowsky, J.~Magueijo, A.~Stebbins, M.~Turner, \& N. Turok
for useful conversations and 
{N. Sugiyama},
R.~Crittenden and A.~de Laix
for isocurvature models.
W.H.~was supported by the NSF and WM Keck Foundation.

\vskip 1.5truecm

\noindent{\tt 

\noindent{\tt whu@sns.ias.edu}

%

\begin{thebibliography}{99}
\bibitem{TiltNote}
Even a tilt of $n=0.6$, for which the first peak height and large-angle 
plateau coincide, gives a very small shift in the peaks and does not obscure
them.  More generally, if significant degree scale power is present
\cite{degree}, the peaks and positions are not obscured 
(see e.g.~\cite{confusion}).
\bibitem{BaryonNote}
If {\it both} the baryon content $\Omega_bh^2\simgt0.03$ and
the CDM content $\Omega_0 h^2\simgt0.6$ are anomalously
high then the second peak (r) will be hidden by the baryon inertia
\cite{BigPaper}.
\bibitem{IsocNote}This does not preclude the possibility of white noise
curvature perturbations at low $k$ generated by pressure perturbations.
\bibitem{TurokNote} We implicitly assume $\rho_\gamma$ is significant
at horizon crossing and $\Phi = -{\cal O}[\Psi]$.
\end{thebibliography}
\end{document}